\documentclass[aps,prl,twocolumn]{revtex4}
\usepackage{amssymb}
\usepackage{amsmath}
\usepackage{amsbsy}
\usepackage{graphicx}
\usepackage{epstopdf}
\graphicspath{{Figures/}}
\usepackage{subfigure}
\newcommand{\CC}{C\nolinebreak\hspace{-.05em}\raisebox{.4ex}{\tiny\bf +}\nolinebreak\hspace{-.10em}\raisebox{.4ex}{\tiny\bf +}}

\begin{document}

\title{A Generic Surface Sampler for Monte Carlo Simulations}
\author{J.\,A.\,Detwiler$^{1, 3}$, R.\,Henning$^{1, 2}$, R.\,A.\,Johnson$^3$, M.\,G.\,Marino$^3$}
\affiliation{$^\textit{1}$\,Lawrence Berkeley National Laboratory, Berkeley, California}
\affiliation{$^\textit{2}$\,University of North Carolina, Chapel Hill, North Carolina}
\affiliation{$^\textit{3}$\,University of Washington, Seattle, Washington}
\date{\today} 

\begin{abstract}
We present an implementation of a Monte Carlo algorithm that generates points 
randomly and uniformly
on a set of arbitrary surfaces. The algorithm is completely general and
only requires the geometry modeling software to provide the intersection
points of an arbitrary line with the surface being sampled. We
demonstrate the algorithm using the Geant4 Monte Carlo simulation
toolkit. The efficiency of the sampling algorithm is discussed,
along with various options in the implementation and example
applications.
\end{abstract}

\maketitle

\section{Introduction}
\label{sec:intro} 

The uniform, random sampling of arbitrarily shaped surfaces  is of
importance in several scientific and technological applications.
For example, generic surface sampling can be used to create and
test more realistic computer graphics models~\cite{Tur92}.
In medical imaging, such sampling can be used to generate a uniform
distribution of target points over the surface of tumors~\cite{Wil87}.
Surface sampling has also been used to study oxygen production in
forests~\cite{Mel04}.
In low-background radiation detection, the application for which
the algorithm presented here was developed, the simulation of
radioactive contaminants on various detector surfaces is important
for quantifying backgrounds and their impact on detector sensitivity.
This algorithm was successfully
implemented into the Geant4-based~\cite{Geant4} simulation toolkit, MaGe~\cite{MaGe},
being jointly developed by the GERDA~\cite{GERDA} and {\sc Majorana}~\cite{MJ} collaborations 
to simulate germanium detector arrays.

Several algorithms exist to perform such generic surface sampling
(see, for example, Refs.~\cite{Tur92} - \cite{Mel04}).
Unfortunately, some of these methods (such as the retiling of
polygonal surfaces) are algorithmically complex and computationally
intensive. Other
algorithms require the surfaces to be represented as differentiable
functions. Deriving such a function for each surface-of-interest
can be a computationally intensive task, particularly for
complex geometries. Finally, to the authors' knowledge, little is available in the form of
free, open-source code for plug-and-play usage.

We have developed a Monte Carlo algorithm that only requires the
geometry modeling software to be able to find
the intersection points between an arbitrary line and the surfaces
of the volumes to be sampled. The algorithm generates a random set of rays that impinge on
the surfaces of interest that are isotropic in direction and uniform
in space.  The intersection points, provided by the geometry modeling software,
are sampled again to provide the final set of random
and uniform surface points.

We demonstrate this generic surface sampling routine using the
\CC-based Geant4 Monte Carlo simulation toolkit~\cite{Geant4}.
Geant 4 is used extensively in high-energy, nuclear and medical
physics to simulate the interactions of radiation with matter. In
Geant4, arbitrary geometries can be constructed by arranging
collections of nested solid volumes and boolean combinations
(intersections, additions, or subtractions) of those volumes in
specified positions and orientations relative to each other. The
available basic solids include fundamental solids such as spheres,
cylinders, and polyhedra, as well as more generic and complex
boundary-representation volumes.  Our sampler relies on the fact
that each Geant4 volume class provides a function that finds the
intersection points between the volume's surface and an arbitrary
line, if such an intersection exists. Each volume class also defines
a function that returns a bounding radius for the volume in question,
which is used to constrain the parameter space of lines sampled.

\section{Sampling Algorithm}
\label{sec:algorithm}

\begin{figure}
\centering
\includegraphics[width=.95\columnwidth]{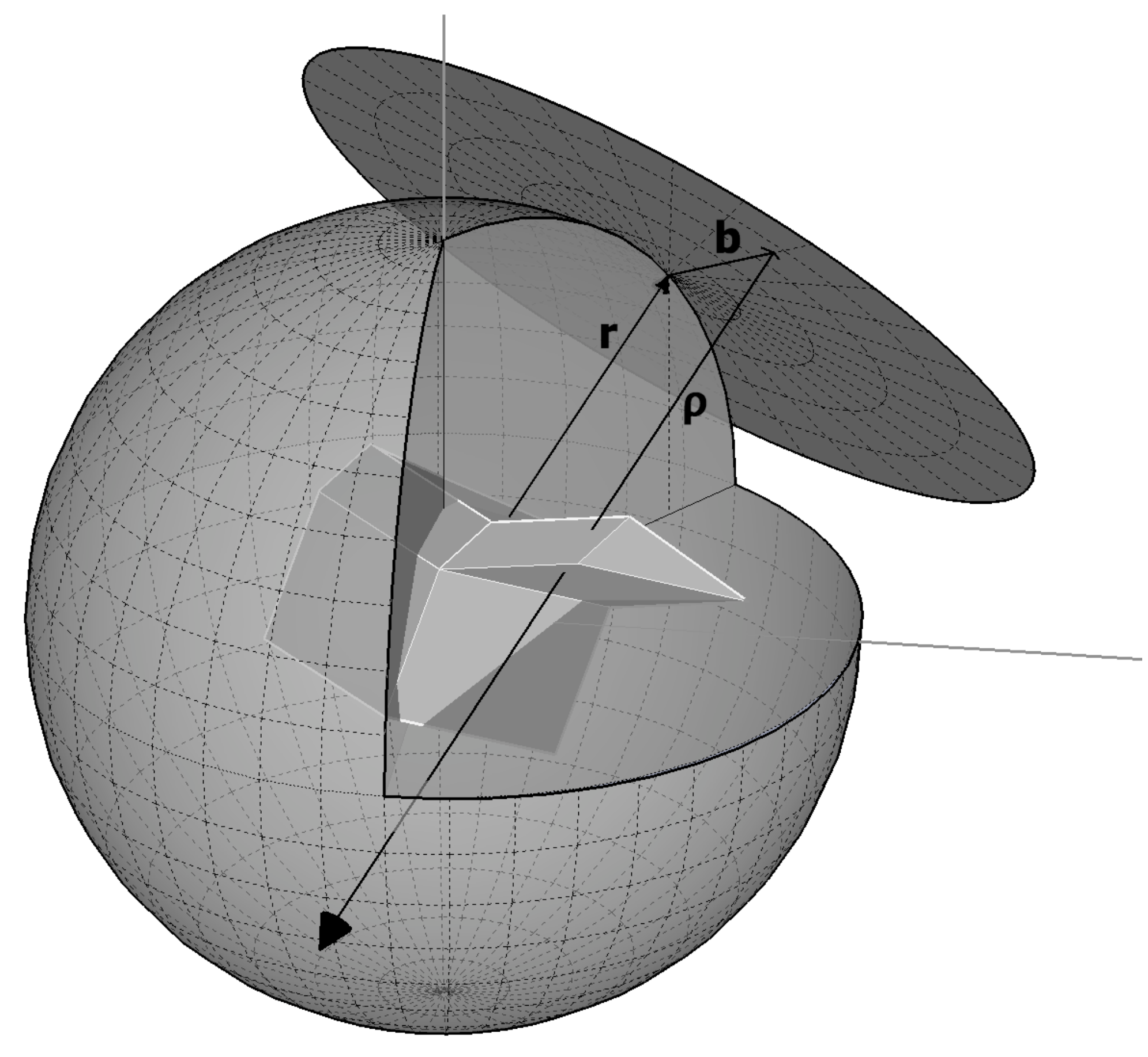}
\caption{An schematic of the bounding sphere 
(shown with a section missing for illustrative purposes),
tangent disk, and ray $\boldsymbol{\rho}$ for a sampling trial of
an arbitrary example volume.  $\boldsymbol{\rho}$ originates
at ${\bf r} + {\bf b}$ and continues in the $-{\bf \hat{\Omega}}$
direction, in this case intersecting the enclosed volume twice.
The determination of the various rays and angles is described in the text.}
\label{fig:rayGeneration}
\end{figure}

The principle of the sampling algorithm is based upon uniformly 
sampling the volume within a sphere.  
When the user selects a volume or set of volumes whose surfaces are
to be sampled, the radius $R$ of a bounding sphere which wholly contains the 
volume(s) must be determined.  In the
case of multiple disjoint volumes, a ``mother'' volume that encompasses
all the volumes to be sampled must be used. In practice, the radius of this bounding
sphere is determined by querying the geometry modeling software.   
To generate a uniform,
isotropic flux of rays within this bounding sphere, first a random
isotropic point ${\bf r}$ on the sphere is generated, where ${\bf r} = R {\bf \hat{\Omega}}$ and 
${\bf \hat{\Omega}}$ is the randomly generated direction.  A disk, also of radius
$R$, is defined tangential to the bounding sphere, with its center
at position ${\bf r}$. Figure~\ref{fig:rayGeneration} shows the position of
this disk and the bounding sphere for a sampling trial of an arbitrary example volume.
The starting point for another ray $\boldsymbol{\rho}$ is
generated on the interior of the disk at point ${\bf r} + {\bf b}$,
where ${\bf b}$ has polar coordinates $(b, \alpha)$ in the coordinate
system of the disk. The ``impact parameter'' $b$ is generated with
a uniform distribution in $b^2$ between 0 and $R$, and the angle $\alpha$ is generated
uniformly between 0 and $2\pi$.  The direction of $\boldsymbol{\rho}$ is taken
to be $-{\bf \hat{\Omega}}$, normal to the circle and hence pointing
into the bounding sphere. The uniformity and isotropy of the
rays produced in this manner will be discussed in the next section.

Once $\boldsymbol{\rho}$  has been generated, the geometry modeling software is
queried to find the intersection points of the ray with all
surfaces among the volumes-of-interest. If no such intersections
exist, another ray is generated with a new direction and starting point. 
If $N$ intersections are found, a
random integer $n$ is generated between one and the maximum number
of intersections possible for the given geometry ($N_{\textrm{max}}$),
which is input by the user~\footnote{A guess will suffice
for the value of $N_{\textrm{max}}$, it merely needs to match or exceed
the greatest number of intersections encountered in the ouput set of 
sampled points. If the algorithm encounters more 
surfaces than $N_{\textrm{max}}$, a warning can be generated and the
user can rerun with a larger value of $N_{\textrm{max}}$.}.  
If $n > N$ the ray is discarded, and the algorithm starts over. 
Otherwise, one of the $N$
intersections is chosen at random.  The set of intersection points
chosen in this way is the output of the algorithm.  

\section{Algorithm Properties}
\label{sec:properties}

In the following, it is assumed that we have a random
number generator that can generate a sequence of real numbers
uniformly distributed between $0$ and $1$, with the standard
requirement of randomness~\cite{Knu81}. Additionally, all vectors,
volumes and surfaces are assumed to lie in 3-dimensional Euclidean
space.

We first show that the flux of rays generated as described above
is uniform and isotropic within the bounding sphere of the
surfaces-of-interest. For every point ${\bf x}$ in the interior of
a sphere of radius $R$, and for every direction ${\bf \hat{\Omega}}$
from ${\bf x}$, there exists one and only one line passing through
${\bf x}$ that is normal to the plane tangent to the sphere at ${\bf
r} = R {\bf \hat{\Omega}}$.  The set of all intersections with this
tangent plane of rays in direction ${\bf \hat{\Omega}}$ originating
from all points ${\bf x}$ interior to the sphere fill a disk of
radius R centered at ${\bf r}$. Since the direction ${\bf \hat{\Omega}}$
is chosen isotropically, and since the starting point on the disk
${\bf b}$ is chosen uniformly across the surface of the disk, then
the probability for a ray to pass within a small area ${\bf \Delta
A}$ centered at ${\bf x}$ with surface normal pointing in direction
${\bf \hat{\Omega}}$ is independent of ${\bf x}$ (uniform), and is
independent of direction ${\bf \hat{\Omega}}$ (isotropic). Symbolically,
we write the normalized vector flux of rays as $\boldsymbol{\phi}({\bf x}, {\bf \hat{\Omega}})
= {\bf \hat{\Omega}} / 4 \pi^2 R^2$, which is independent
of ${\bf x}$.  The randomness of this flux
is guaranteed as long as a new direction ${\bf \hat{\Omega}}$ and
disk position ${\bf b}$ are chosen for each ray.

The uniformity and randomness of the set of intersection points
generated by the uniform isotropic flux of rays can be demonstrated
as follows.  First, divide the surfaces-of-interest into an large number
of surface elements ${\bf \Delta A}({\bf x})$, where the direction
points normal to the surface at point ${\bf x}$, and the magnitude
$\Delta A$ is independent of ${\bf x}$ (${\bf \Delta A}({\bf x}) =
\Delta A {\bf \hat{n}}({\bf x})$). $\Delta A$ is taken to be small enough
that each surface element may be approximated to be flat \footnote{This
is equivalent to requiring that the sampled surfaces be differentiable.
Within the Geant4 framework, this implies a requirement that the
radius-of-curvature for any surface be much greater than the tolerance
parameter, which sets the distance within which a point is considered
to be ``on'' a volume's surface. This parameter has a default value
of 1 pm, but can be tuned by the user to be as low as $\sim$1 fm
for typical meter-sized or smaller geometries, at which point one
is limited by numerical round-off of the 64-bit double-precision
floating point data type used to define volume dimensions. The
assumption of flatness of the surface elements also neglects
infinitely sharp corners, which are unphysical.}.  The
probability for a surface element to be hit by a ray from our
generated vector flux $\boldsymbol{\phi}({\bf x}, {\bf \hat{\Omega}})$ is
\begin{eqnarray*}
\int_0^{4\pi} \left|\boldsymbol{\phi}({\bf x}, \Omega) \cdot {\bf \Delta A}({\bf x})\right|~d\Omega & = & \int_0^{4\pi} \left|\frac{\Delta A}{4 \pi^2 R^2} ~{\bf \hat{\Omega}} \cdot {\bf \hat{n}}({\bf x})\right|~d\Omega \\ 
& = & \frac{\Delta A}{4 \pi^2 R^2} \int_0^{4\pi} \left|\cos\theta\right|~d\Omega\\ 
& = & \frac{\Delta A}{2 \pi R^2} \\
\end{eqnarray*}
which is independent of ${\bf x}$. This implies that all surface
elements are hit with constant probability. Thus the set of
intersections of all rays with the surfaces-of-interest gives a
uniform sampling of those surfaces.

The randomness of initial flux of rays implies that the set of
intersection points generated by one ray is statistically independent
from those of other rays. However, intersection points of a single
ray are not statistically independent from each other, as they all
lie along a single line. For a truly random sampling, at most one
intersection point can be chosen from each ray.  Note that if a
single point were chosen at random and kept for each ray with
intersections, those points which lie along rays with fewer
intersections would be sampled more often than those points lying
along lines with more intersections, ruining the uniformity of the
distribution.  In essence, rays with $N$ intersections would effectively
be given a $1/N$ weighting.
For this reason, the point selection is weighted by
$N/N_{\textrm{max}}$, and uniformity is retained. 

The efficiency of the above method, in terms of the number of surface
points generated per geometrical calculation, can be poor when the
volumes-of-interest sparsely fill the bounding sphere. If the volumes
are disjoint, efficiency can be recovered by considering distant
volumes independently. Poor efficiency for volumes having needle-like
or planar geometries, with one dimension much larger or smaller
than the other dimensions, can be remedied by considering bounding
surface other than a sphere, for example a wide plane or a narrow
cylinder.  In such cases care must be taken to ensure the generated
flux of rays is (at least approximately) uniform and isoptropic.
We did not consider such cases in this paper.

The step in which rays with fewer intersections are preferentially
discarded also imposes an efficiency reduction by a factor of roughly
$\bar{N}/N_{\textrm{max}}$, where $\bar{N}$ is the average number
of intersections per ray. This reduction can be significant for
geometries with many aligned, repeated volumes, as well as for
geometries with regions containing many small components. In such
cases it may be prudent to simply keep all intersection points of
all rays. The resulting set of points, taken as a whole, will still
distribute with uniform surface density, and with much higher
efficiency, albeit at the cost of introducing correlations among
some consecutive points. For many applications, though, such correlations
are irrelevant.

\section{Geant4 Implementation}

We implemented this algorithm within the Geant4 framework by deriving
classes from the ``user action'' base classes \emph{G4VUserPrimaryGeneratorAction} and
\emph{G4UserSteppingAction}.  At runtime the user
inputs a list of volume names whose surfaces are to be sampled,
which are sent to the generator action class.  After geometry
initialization, the class queries the \emph{G4PhysicalVolumeStore}
to find the smallest volume which contains all volumes-of-interest
as daughter volumes (this volume may itself be a volume-of-interest).
The \emph{G4VSolid} corresponding to that mother volume is extracted
from its \emph{G4LogicalVolume}.  A bounding radius for the
surfaces-of-interest is then obtained by calling
\begin{verbatim}
G4VSolid::GetExtent().GetExtentRadius();
\end{verbatim}
The class then sets the primary particle to be a ``geantino'', an
imaginary neutral, massless utility ``particle" within the Geant4
framework which undergoes no interactions, and only travels in
straight lines. Geantinos are commonly used for debugging purposes
and to map out geometries.  The geantino's position and direction
are selected by our algorithm to give a uniform, isotropic flux of
geantinos throughout the interior of the bounding sphere. The energy
of the geantino can be any value greater than 0.  The choice of
geantinos delegates all geometrical calculations to Geant4.

The stepping action class checks at each step whether the geantino
is entering or exiting a volume of interest. Each such entrance or
exit point is added to a list of surface intersections. At the end
of the event, one of these surface intersections can be chosen at
random, or all surface
intersections can be kept if efficiency requirements outweigh
the necessity for truly uncorrelated sampling. The set of surface
intersections generated in this way uniformly sample the surfaces
of interest, and may be saved to disk or used for further processing
in the program (e.g.~as the vertex for the next event).

\section{Example Application and Verification}

Such an implementation of our generic surface sampling algorithm
was added to MaGe~\cite{MaGe}, a Geant4-based Monte Carlo simulation
toolkit optimized for low-background germanium detector simulations.
The output vertices are written to a ROOT~\cite{ROOT} file, which can then be
used in simulations involving surface physics, for example
$\alpha$-particle backgrounds from natural U and Th decay chain isotopes in settled dust,
or from Rn decay chain daughters plated out on detector surfaces.

\begin{figure*}
\centering
  \subfigure[~Rendering of a 3-Ge-crystal ``string'' assembly. The entire assembly is about 30~cm in length.]{\label{fig:StringRender} \includegraphics[width=.8\columnwidth]{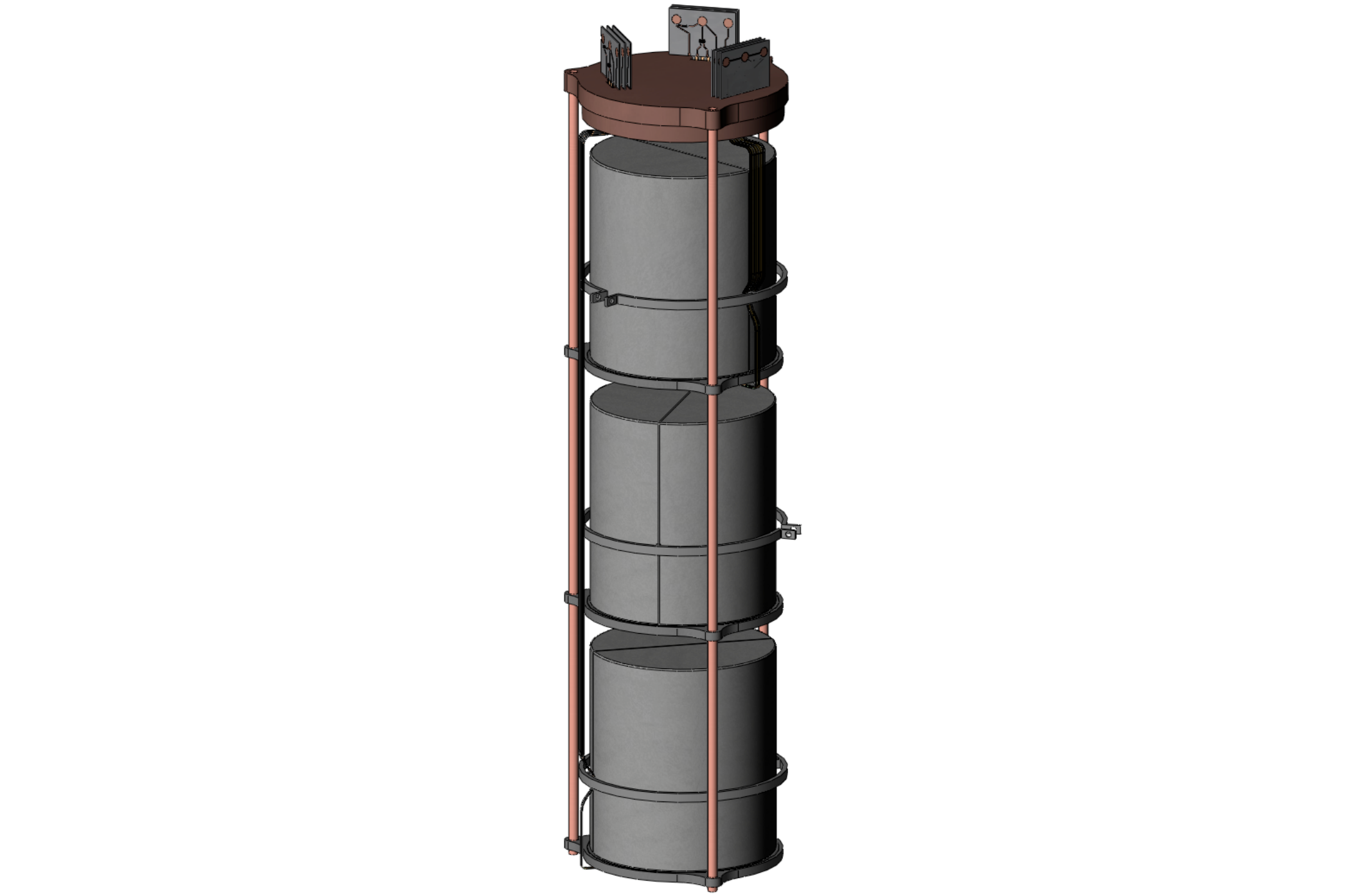}}
  \subfigure[~Horizontal view of 19 strings hanging from a Cu coldplate, imaged with our surface sampling algorithm.]{\label{fig:57BangerSamp} \includegraphics[width=.8\columnwidth]{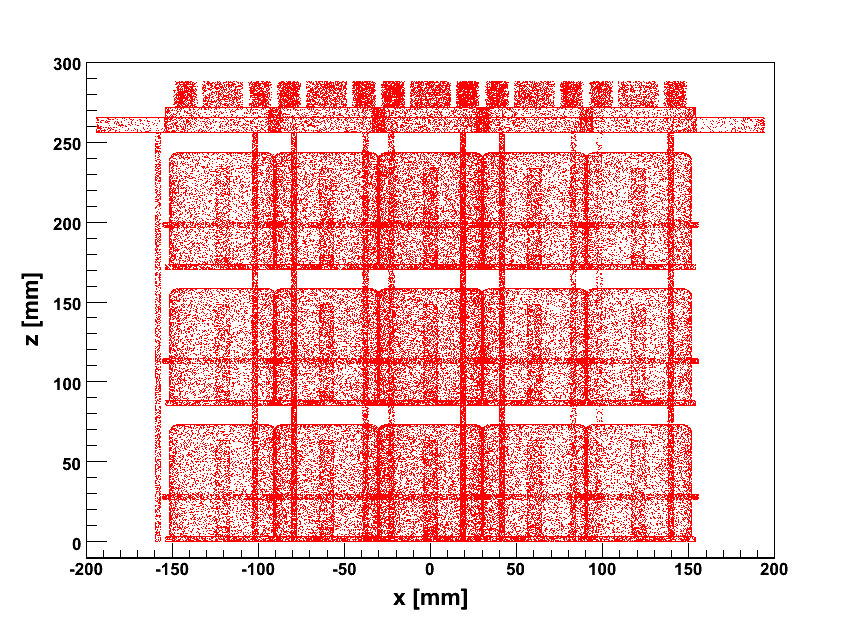}}
  \subfigure[~Surface sampling of a close-ended coaxial high-purity germanium detector crystal.]{\label{fig:SingXtalCoax} \includegraphics[width=.8\columnwidth]{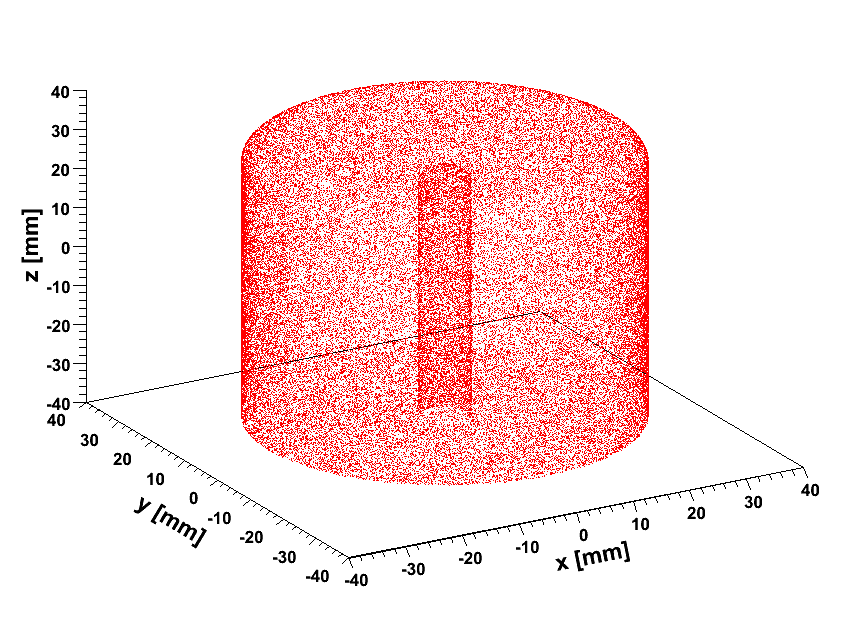}}
  \subfigure[~Surface sampling of a crystal support tray. A rendering of the simulated geometry is shown in the upper right corner.]{\label{fig:TraysSamp} \includegraphics[width=.8\columnwidth]{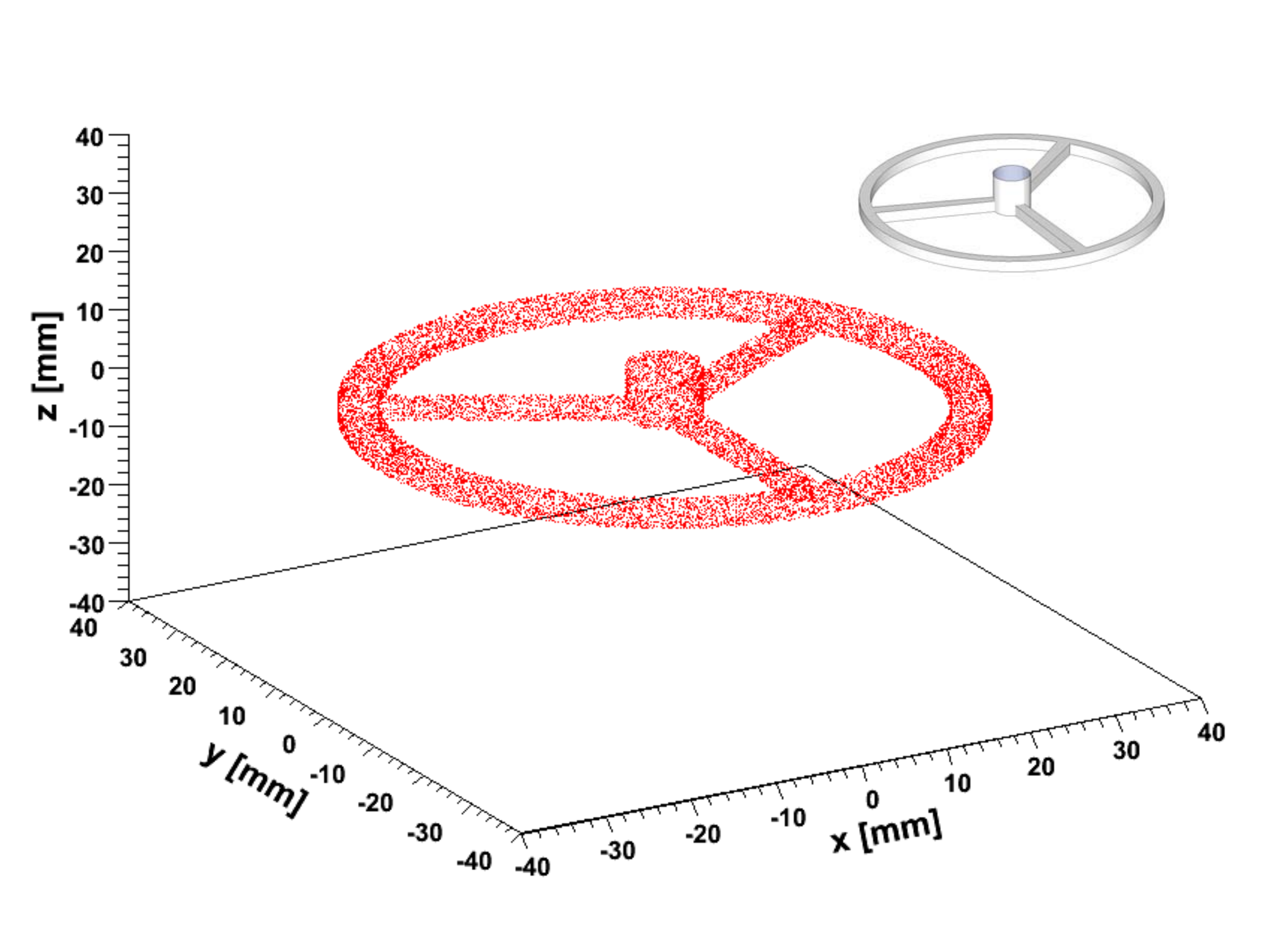}} 
  \caption{Demonstration of the uniform surface sampling on various
  volumes in the {\sc Majorana} 57-crystal array design.  The
  2-dimensional projection of 3-dimensional points leads to regions
  with apparent higher or lower sampling densities, for example at
  the edges of the displayed geometries.  See Table~\ref{tab:Verification}
  for an analytic verification of the sampler. }
  \label{fig:demonstration}
\end{figure*}

Figure~\ref{fig:demonstration} demonstrates the usage of the surface sampler 
on the 57-detector array design for the {\sc Majorana} experiment~\cite{MJ}.
Figure~\ref{fig:StringRender} shows a rendering of a 3-Ge-crystal string assembly,
complete with detector supports and electronic connections and components. 19
such strings are arranged in a hexagonal close-pack pattern, suspended from a 
Cu cold plate, and housed in a cylindrical low-background cryostat made of 
electroformed Cu. An imaging of the 
full simulated geometry (minus the surrounding cryostat) 
by our surface sampling algorithm is shown in 
Figure~\ref{fig:57BangerSamp}, as viewed from one side. 
We also show more detailed samplings of two
specific detector components in Figures~\ref{fig:SingXtalCoax} and \ref{fig:TraysSamp}.
Figure~\ref{fig:SingXtalCoax} plots the output of the algorithm for
one close-ended coaxial high-purity germanium detector crystal. The detector is represented
by a boolean combination of basic volumes. The body is modeled as
two cylinders OR'd with a torus to form the rounded top face. A
third, smaller-radius cylinder OR'd with a sphere at one end is
subtracted from the body to form the coaxial well along the detector's
vertical axis.  Figure~\ref{fig:TraysSamp} shows a surface sampling of one
of the plastic trays on which the Ge crystals rest in the string assembly. A
rendering of the simulated tray design is shown to the upper right of the surface 
sampling for comparison.

We ran a high statistics simulation to test the behavior of the surface sampler
and verify that the surface density of sampled points is
independent of surface shape and orientation.  To this end, we sampled a portion of the
{\sc Majorana} 57-detector array design.  We chose to 
sample the inner surface of the enclosing
cylindrical cryostat, the cold plate from which all the crystals hang, two
crystal detectors, and a single ``contact ring" (a thin plastic ring 
that clamps leads against the crystal surface to make electrical connections
to the detector) surrounding one of the crystals.
The inner cryostat surface and the cold plate are both simple cylinders.
The contact ring is an annulus, and the detectors are as described
above. All surfaces were sampled simultaneously, so the surface density
of sampled points should be the same for all five components.
The ratio of points on a volume's surface to total number of sampled
points in the run were tabulated from the output ROOT file.  These
ratios were then compared with analytical calculations of the surface
area ratio for each volume to the total surface area of all sampled
volumes.  The results are shown in
Table~\ref{tab:Verification}. In all cases, the ratios agree within the
sampled statistics.

\begin{table}
\caption{A comparison of analytically calculated surface area ratios
to the fractions of sampled points landing on each surface of a
number of volumes sampled simulataneously using our generic surface
sampling algorithm. In all cases, the ratios agree within the
statistics of the simulation.}
\label{tab:Verification}
\begin{center}
\begin{tabular}{|l|r@{.}l|r@{.}l|}
\hline
 &  \multicolumn{2}{|c|}{Analytic [\%]} & \multicolumn{2}{|c|}{Sampled [\%]} \\ \hline
Cryostat        & ~~~~69  & 544  & ~69 & 577 $\pm$ 0.042~ \\
Cold Plate      & 25  & 906  & 25 & 881 $\pm$ 0.026 \\
Detector 1      & 2   & 173  & 2  & 171 $\pm$ 0.007 \\
Detector 2      & 2   & 173  & 2  & 167 $\pm$ 0.007 \\
Contact Ring    & 0   & 202  & 0  & 203 $\pm$ 0.002 \\ \hline
\end{tabular}
\end{center}
\end{table}

\section{Concluding Remarks}

We have developed a generic surface sampling algorithm that distributes
vertices uniformly and randomly over sets of arbitrary surfaces.
Such an algorithm has potential application in many scientific and
technical fields. Our implementation within the Geant4 Monte Carlo
simulation toolkit and the MaGe simulation framework for germanium
detector-based systems is of particular use to nuclear and particle
physicists. It may be used, for example, to study surface $\alpha$
backgrounds, a key background in many low-background calorimetry-based
experiments in these fields.

\section{Acknowledgments}

This work was sponsored in part by the US Department of
Energy under Grant nos. DE-FG02-97ER41020 and
DE-AC02-05CH11231. This research used the Parallel
Distributed Systems Facility at the National Energy
Research Scientific Computing Center, which is supported
by the Office of Science of the U.S. Department of Energy
under Contract no. DE-AC02-05CH11231. The authors would 
like to thank D.\,Y.\,Sebe for assistance with some of the 
figures.  In addition, the authors acknowledge the MaGe group of the
GERDA and {\sc Majorana} collaborations for important comments
and insight.

\end{document}